%% file: main.tex
\documentclass[twocolumn]{aastex631}

\usepackage{comment}
\usepackage{graphicx}
\usepackage{gensymb}

\newcommand{\lowhi}[2]{$_{-#1}^{+#2}$}
\newcommand{\percent}[0]{\SI{}{\percent}}

\newcommand{\RNum}[1]{\uppercase\expandafter{\romannumeral #1\relax}}

\newcommand{\radrat}[0]{R\textsubscript{p}/R\textsubscript{$\star$}}
\newcommand{\rp}[0]{R\textsubscript{p}}
\newcommand{\rs}[0]{R\textsubscript{$\star$}}
\newcommand{\rj}[0]{R\textsubscript{Jup.}}

\newcommand{\sr}[0]{\textsubscript{$\star$}}
\newcommand{\mpj}[0]{M\textsubscript{p}}
\newcommand{\mj}[0]{M\textsubscript{Jup.}}

\begin{document}

\title{Colorado Ultraviolet Transit Experiment Near-Ultraviolet Transmission Spectroscopy of the Ultra-hot Jupiter KELT-9b}

\author[0000-0002-4701-8916]{Arika Egan}
\affiliation{Laboratory for Atmospheric and Space Physics, University of Colorado Boulder, 1234 Innovation Drive, Boulder, CO 80303}
\affiliation{Applied Physics Laboratory, Johns Hopkins University 11101 Johns Hopkins Rd, Laurel, MD 20723 }

\author[0000-0002-1002-3674]{Kevin France}
\affiliation{Laboratory for Atmospheric and Space Physics, University of Colorado Boulder, 1234 Innovation Drive, Boulder, CO 80303}

\author[0000-0002-4166-4263]{Aickara Gopinathan Sreejith}
\affiliation{Space Research Institute, Austrian Academy of Sciences, Schmiedlstrasse 6, A-8042, Graz, Austria}
\affiliation{Laboratory for Atmospheric and Space Physics, University of Colorado Boulder, 1234 Innovation Drive, Boulder, CO 80303}


\author[0000-0003-4426-9530]{Luca Fossati}
\affiliation{Space Research Institute, Austrian Academy of Sciences, Schmiedlstrasse 6, A-8042, Graz, Austria}

\author[0000-0003-3071-8358]{Tommi Koskinen}
\affiliation{Lunar and Planetary Laboratory, University of Arizona, Tucson, AZ 85721, USA}

\author[0000-0002-2129-0292]{Brian Fleming}
\affiliation{Laboratory for Atmospheric and Space Physics, University of Colorado Boulder, Boulder, CO 80303, USA}

\author[0000-0001-7131-7978]{Nicholas Nell}
\affiliation{Laboratory for Atmospheric and Space Physics, University of Colorado Boulder, Boulder, CO 80303, USA}

\author[0000-0002-0506-0825]{Ambily Suresh}
\affiliation{Laboratory for Atmospheric and Space Physics, University of Colorado Boulder, 1234 Innovation Drive, Boulder, CO 80303}

\author[0000-0001-9207-0564]{P. Wilson Cauley}
\affiliation{Laboratory for Atmospheric and Space Physics, University of Colorado Boulder, 1234 Innovation Drive, Boulder, CO 80303}

\author[0000-0002-0875-8401]{Jean-Michel Desert}
\affiliation{Anton Pannekoek Institute of Astronomy, University of Amsterdam, Amsterdam, The Netherlands}

\author[0000-0001-7624-9222]{Pascal Petit}
\affiliation{Institut de Recherche en Astrophysique et Planétologie, Université de Toulouse, CNRS, CNES, 14 avenue Edouard Belin, F-31400 Toulouse, France}

\author[0000-0001-5371-2675]{Aline A. Vidotto}
\affiliation{Leiden Observatory, Leiden University, P.O. Box 9513, 2300 RA, Leiden, The Netherlands}

\begin{abstract}
We present new near-ultraviolet (NUV, $\lambda$ = 2479 -- 3306 \AA) transmission spectroscopy of KELT-9b, the hottest known exoplanet, obtained with the Colorado Ultraviolet Transit Experiment ($CUTE$) CubeSat. Two transits were observed on September 28th and September 29th 2022, referred to as Visits 1 and 2 respectively. Using a combined transit and systematics model for each visit, the best-fit broadband NUV light curves are \radrat $=$ 0.136\lowhi{0.0146}{0.0125} for Visit 1 and \radrat $=$ 0.111\lowhi{0.0190}{0.0162} for Visit 2, appearing an average of 1.54$\times$ larger in the NUV than at optical wavelengths. While the systematics between the two visits vary considerably, the two broadband NUV light curves are consistent with each other. A transmission spectrum with 25 \AA\ bins suggests a general trend of excess absorption in the NUV, consistent with expectations for ultra-hot Jupiters. Although we see an extended atmosphere in the NUV, the reduced data lack the sensitivity to probe individual spectral lines.






\end{abstract}


\input{introduction}

\input{observations}

\input{dataReduction}

\input{results}

\input{conclusion}

\begin{acknowledgments}
\section*{acknowledgments}
Much of this work was funded by NASA grants NNX17AI84G and 80NSSC21K166 (PI- K. France). AAV acknowledges funding from the European Research Council (ERC) under the European Union's Horizon 2020 research and innovation programme (grant agreement No 817540, ASTROFLOW). A.G.S. was supported by the Schrödinger Fellowship through the Austrian Science Fund (FWF) [J 4596-N]. We are additionally grateful for the thoughtful and thorough feedback provided by the reviewer.
\end{acknowledgments}

%

\vspace{5mm}
\facilities{$CUTE$}


\software{\texttt{batman} \citep{Kreidberg2015_batman}, 
          \texttt{lmfit} \cite{lmfit},
          \texttt{CONTROL} \cite{Sreejith_2022},
          \texttt{scikit-learn} \cite{scikit-learn}
          }

\appendix

\input{appendix_lmfit}

\bibliography{bibtex}
\bibliographystyle{aasjournal}

\end{document}

%% file: introduction.tex
\section{Introduction}
 KELT-9b is the hottest exoplanet discovered to date with an equilibrium temperature  T\textsubscript{eq} $\simeq$ 3921 K \citep{Borsa2019}. The planet has \rp\ = 1.783 $\pm$ 0.009 \rj , mass \mpj\ = 2.44 $\pm$ 0.70 \mj\ \citep{Hoeijmakers2019}, and orbits an A0 star with T\textsubscript{eff} = 9495 $\pm$ 104 K \citep{Kama2023} at a distance $a$ = 0.0336 AU \citep{Kama2023} with a period $P$ = 1.4811 days \citep{Ivshina2022}, an environment which prevents aerosol formation on the planet's day side \citep{Lothringer2018_models, Kitzmann2018} and possibly drives an escaping outflow. Both H$\alpha$ \citep{Yan2018_Halph_kelt9, Cauley2019} and H$\beta$ \citep{Cauley2019} were detected to be optically thick out to $\sim$1.61 \rp. Several metals have also been detected in the planet's atmosphere (out to 1.10 \rp\ and between 10$^{-3}$ and 10$^{-6}$ bar) with ground-based high-resolution transmission spectroscopy, including \ion{Fe}{1}, \ion{Fe}{2}, \ion{Ti}{2}, \ion{Mg}{1}, \ion{Na}{1}, \ion{Cr}{1}, \ion{Sc}{2}, and \ion{Ca}{2} \citep{Hoeijmakers2018, Hoeijmakers2019, Lowson2023, Cauley2019, Borsato2023FOCES, Turner2020, Yan2019_CaII, Borsato2023Mantis, BelloArufe2022}. The \ion{O}{1} triplet at 7774 \AA\ was measured by \cite{Borsa2022}, and most recently, \cite{Borsato2023Mantis} reported new detections of \ion{Ni}{1}, \ion{Sr}{1}, \ion{Tb}{2}, \ion{V}{1}, and \ion{Ba}{2}. 

 
These ground-based detections have provided important constraints on the abundances, temperature-pressure (TP) profile, and overall structure of the atmosphere. Both H-Balmer lines and metal lines are best fit with model atmospheres that include non-local thermodynamic effects (NLTE) \citep{garciamunoz2019, Turner2020, Borsa2022}. Compared with LTE models, the NLTE models predict upper atmospheric temperatures on the order of 8,000 K \citep{Fossati2021, Turner2020, Borsa2022}.

Quantities of the upper atmosphere, like abundances and mass-loss estimates, have been made using optical absorption features, but these observations are unable to directly probe the upper atmosphere \citep{Fossati2021}. For example, the H$\alpha$ and \ion{Ca}{2} detections from \cite{Turner2020} were found at effective altitudes between 1.2 $-$ 1.44 \rp. The \ion{O}{1} triplet in \cite{Borsa2022} was measured with an effective altitude of 1.17 \rp. KELT-9b is believed to have an escaping atmosphere, but detections of such have not been conclusively made, as the planet's Roche lobe is at 2.017 \rp\ (Section \ref{sec:results}).

 
Ultraviolet transmission spectroscopy can provide a unique complement to atmospheric characterization studies. The near-ultraviolet (NUV) bandpass between $\sim$ 1800 $-$ 3500 \AA\ contains hundreds of strong and abundant metal lines like Fe and Mg \citep{Fossati2010, Haswell2012} that have been observed escaping in several planetary atmospheres \citep{Sing2019, Sreejith2023}. Observations of high-altitude and escaping metal lines can constrain the energy balance and ionization states of the upper atmospheric layers \citep{Cubillos2020, Cubillos2023}. For example, Mg  important coolant in the upper atmospheres of HD189733b \citep{Huang2017} and KELT-20b \citep{Fossati2023}, and Fe was found to be strongly tied to atmospheric heating \citep{Fossati2021}. In addition, NUV observations have been used to assess the presence of scattering hazes \citep{Lothringer2022UV, Wakeford2020, Gressier2023}.

Here we present new NUV observations of KELT-9b obtained with the Colorado Ultraviolet Transit Experiment ($CUTE$), a CubeSat mission dedicated to exploring the upper atmospheres of highly-irradiated ultra-hot Jupiters \citep{France2023}. We first provide an overview of the $CUTE$ CubeSat and describe the KELT-9b observations in Section \ref{sec:observations}. Section \ref{sec:dataAnalysis} describes the data reduction and general light curve modeling. Broadband NUV light curves with 100 \AA\ and 25 \AA\ transmission spectra are presented in Section \ref{sec:results}, and a summary is provided in Section \ref{sec:conclusion}.

%% file: observations.tex
\section{Observatory \& Observations}\label{sec:observations}
\subsection{$CUTE$ Instrument Description}\label{subsec:observatory}
\begin{figure*}
    \centering
    \includegraphics[width = \linewidth]{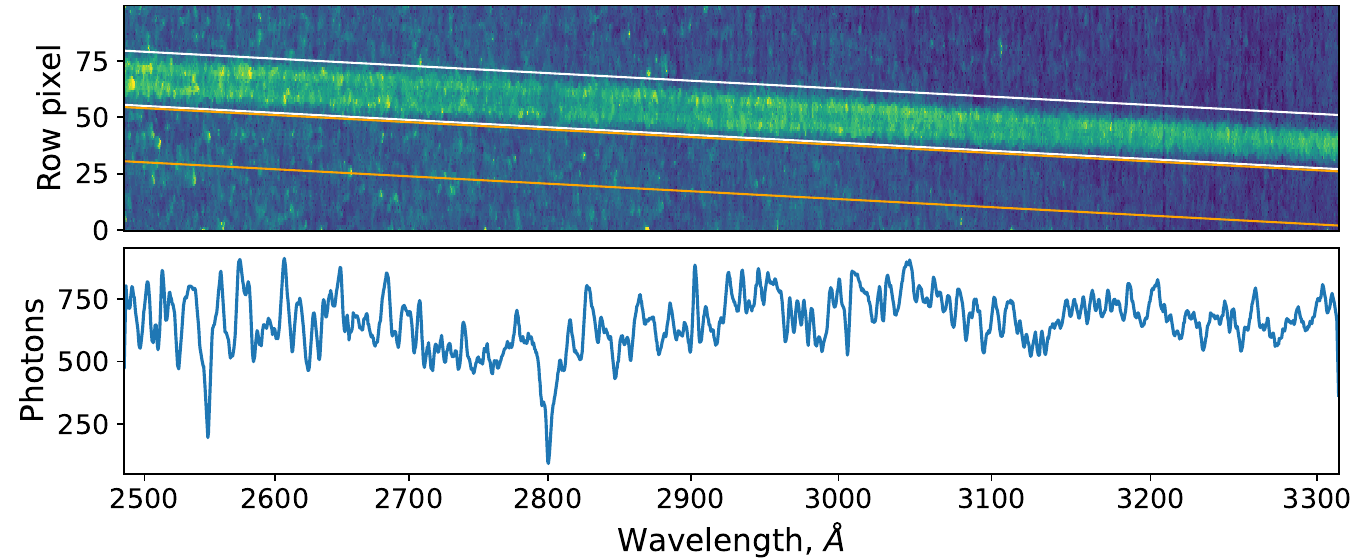}
    \caption{\textbf{Top:} A 2D $CUTE$ CCD image of a single 5-minute exposure. The image is bias-subtracted and corrected for bad pixels and cosmic rays. The spectral extraction region is outlined in white, and the background region is outlined in orange. \textbf{Bottom:} The summed 1D spectrum after background subtraction, smoothed with a Gaussian filter for display purposes.}
    \label{fig:1d2d}
\end{figure*}

The $CUTE$ instrument is an NUV spectrograph operating between 2479 \AA\ and 3306 \AA\ with R $\sim$ 750 and an average dispersion of 0.404 \AA\ pixel\textsuperscript{-1}. $CUTE$ operates in low-Earth orbit with a period of about 95 minutes and an inclination of $\sim$98$\degree$. Like transit observations obtained with the \textit{Hubble Space Telescope} ($HST$), $CUTE$ time-series observations exhibit gaps in coverage due to Earth occultations. The spectrum is recorded on a passively-cooled back-illuminated CCD with 515 $\times$ 2048 (spatial $\times$ spectral) active pixels \citep{Nell2021}. The CCD experiences temperatures between $-$5 $\degree$C and $-$12 $\degree$C over the course of an orbit, resulting in a temperature-dependent dark current that is included in our systematics modelling and removal, as described in Section \ref{sec:dataAnalysis} and \cite{Egan2023}. A two-dimensional spectrum trimmed to 100 rows centered around the spectral trace, and the one-dimensional spectrum, are shown in Figure \ref{fig:1d2d}.




%


\input{obs_table}

\subsection{KELT-9b observations}\label{subsec:observations}
Two successive transits of KELT-9b were observed with $CUTE$ in 2022 on September 28th and September 29th, referred to hereafter as Visit 1 and Visit 2, respectively. Each visit was planned such that the full observation window was 5$\times$ the 3.91 hour transit duration, resulting in $\approx$ $\pm$ 10.5 hours on either side of the transit center. Bias and dark frames were obtained before and after each observation window with a 0.75$\degree$ pointing offset to point the $CUTE$ aperture at dark sky but maintain similar sky background levels. Exposures are 5 minutes each and Visit 1 consists of 13 $CUTE$ orbits and Visit 2 consists of 15 orbits, each with 4 CCD exposures per orbit. While the $CUTE$ spacecraft is in Earth's shadow for approximately half of the 95-minute orbit, Earth and moon avoidance angles as well as time required for the spacecraft to settle into fine-pointing mode reduce the total available observing time to approximately 20 -- 30 minutes per orbit. Occasionally, the $CUTE$ spacecraft exhibits anomalously high pointing jitter within a given 5 minute observation that smears the spectrum across a larger region of the detector. This reduces the per-pixel signal to noise, and integrating over more pixels results in a reduction of the overall signal-to-noise ratio (SNR) of those exposures compared to a low-jitter frame, rendering them unusable in light curve analysis. Observations with a jitter $>$ 10\arcsec\ RMS are excluded from light curve analysis. Table \ref{tab:kelt9b_obs} provides a summary of the observations analyzed herein.

%% file: obs_table.tex
\begin{deluxetable*}{ccccccc} 




\tablecaption{CUTE KELT-9b Observations\label{tab:kelt9b_obs}}


\tablehead{\colhead{Visit \#} & \colhead{Visit Start} & \colhead{Mid-transit Time}& \colhead{Visit End} & \colhead{Total Orbits} & \colhead{Total Frames} & \colhead{Valid Frames\tablenotemark{a}} \\ 
\colhead{} & \colhead{(2022 UTC)} & \colhead{(2022 UTC)}& \colhead{(2022 UTC)}  & \colhead{} & \colhead{} & \colhead{}}

\startdata
1 &    Sept. 27 - 14:54 & Sept. 28 - 01:36  & Sept. 28 - 12:02 &    13 &    46 & 44\\
2 &    Sept. 29 - 02:09 & Sept. 29 -  13:09 & Sept. 30 - 00:47 &    15 &    57 & 54\\
\enddata

\tablenotetext{a}{A valid frame has spacecraft pointing jitter $<$ 10\arcsec\ RMS.}

\end{deluxetable*}

%% file: dataReduction.tex
\section{Data Analysis}\label{sec:dataAnalysis}
\subsection{Data Reduction}\label{subsec:reduction}
The data were reduced using a modified process of the $CUTE$ Data Reduction Pipeline \citep{Sreejith_2022}. Raw stellar frames are corrected for bad pixels and cosmic rays. Bias frames are corrected for bad pixels and cosmic rays, and a master bias frame is created by taking the median of the corrected individual bias frames from each visit; this master bias frame is then subtracted from all stellar frames. The background is estimated and subtracted using the bias-subtracted frame. Bad pixels are replaced with the median of a 3$\times$3 grid of non-bad pixels surrounding the bad pixel. Cosmic rays are identified and replaced using the L.A.Cosmic routine \citep{vanDokkum2001}. A region is defined around the spectrum and summed to create a one-dimensional stellar $+$ background spectrum. A background region below the spectrum of the same size as the stellar region is summed to form a background one-dimensional spectrum that is subtracted from the stellar $+$ background spectrum to produce the final 1D stellar spectrum. The stellar and background regions are shown in the top panel of Figure \ref{fig:1d2d} as surrounded by the white and orange lines respectively, and the corresponding 1D stellar spectrum is shown in the bottom panel of Figure \ref{fig:1d2d}. To create a broadband NUV light curve, the entire spectral bandpass is summed and plotted in time; the NUV light curves for Visits 1 and 2 are shown in Figure \ref{fig:white_light_v1v2}.

\subsection{Light curve \& systematics fitting}
\begin{figure}
    \centering
        \includegraphics[width=0.47\textwidth, trim={0.40cm 0 0 0},clip]{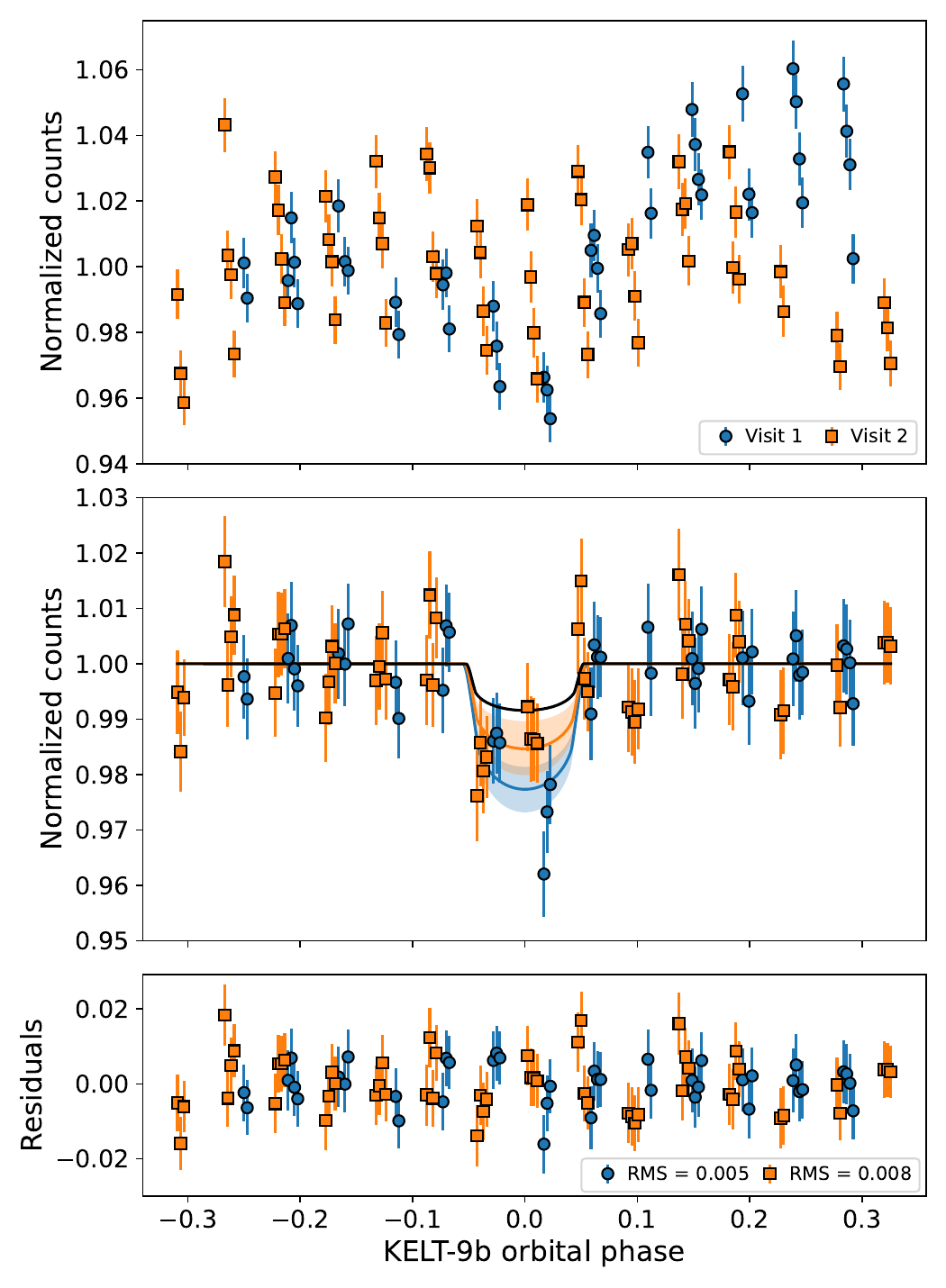}
     \caption{$CUTE$ raw, broadband NUV light curve best fits, and residuals for KELT-9b Visit 1 (blue circles) and Visit 2 (orange squares). Each data point represents a 5-minute exposure. \textbf{Top: }The uncorrected counts normalized to the median of the first three $CUTE$ orbits, which are out of transit. \textbf{Middle: }The normalized counts with the best-fit instrument systematics model removed, 1$\sigma$ error bars and each visit's light curve model in their respective colors. The 1$\sigma$ \radrat\ regions are transparently shown in each visit's respective colors. The black line transit is the optical transit light curve from \cite{Wong2020}. \textbf{Bottom: }The residuals with 1$\sigma$ error bars. The two light curves are shown on the same plot to emphasize the difference in systematics across each visit.}\label{fig:white_light_v1v2}
\end{figure}

Like other space telescopes (e.g. $HST$; \citealt{Wakeford2016, Gressier2023}) $CUTE$ exhibits orbital- and visit-dependent systematics. As seen in the top plot of Figure \ref{fig:white_light_v1v2}, the systematics embedded in both visits take different forms, though a transit signal appears to be present in both. Out of transit, the uncorrected counts vary by 10.8\% in Visit 1 and 8.8\% in Visit 2. $CUTE$ systematics are heavily correlated with its orbit, evident in the trend of decreasing counts as a function of $CUTE$ orbit, a trend visible in the normalized uncorrected light curves for both Visits 1 and 2. These trends are strongly related to the temperature of the CCD, which varies approximately 6 $\degree$C per orbit \citep{Egan2023}, starting at higher temperatures as $CUTE$ enters the Earth's shadow and cools until exiting the shadow. In addition to thermal trends, $CUTE$'s wide field of view and compact design (\citealt{Fleming2018, egan2020}) subjects the focal plane to low scattered light levels that change throughout an orbit and throughout a visit. Finally, pointing jitter increases the extent of the spectral trace on the detector, reducing the observation's signal-to-noise ratio.

There are several parameters that describe $CUTE$'s pointing that are related to the CCD temperature, scattered light levels, and exhibit covariances amongst each other (e.g. Figure 3 in \citealt{Egan2023}): Azimuth and elevation angles of the telescope with respect to (1) the Earth, (2) the Sun, and (3) the Moon; the Earth latitude and longitude over which the observation began; the CCD temperature; and the R.A., Dec., and roll angles of the telescope. Several of these 12 parameters are necessary to detrend the raw light curves and isolate the astrophysical signal from the background. Individually, all of these parameters are correlated with the background levels by some function (e.g. linear, quadratically), as well as related to each other. Due to the covariance among the parameters, it becomes challenging to include all of them, or select a subset that best describes the systematics. To include as many parameters as possible in the detrending analysis, we utilized principal component analysis (PCA) using the 12 parameters listed above.

\begin{figure*}[t!]
    \gridline{\fig{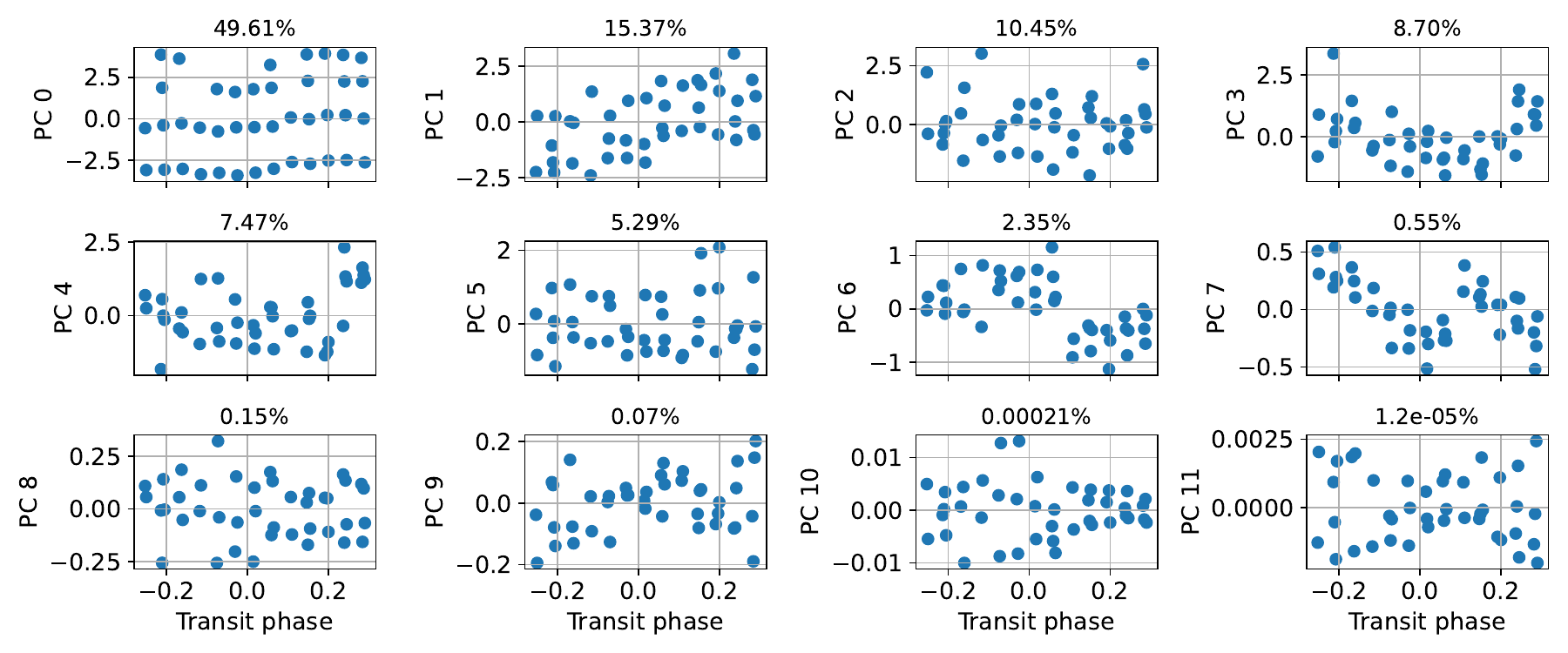}{0.95\textwidth}{}}
    \vspace{-0.8cm}
    \gridline{\fig{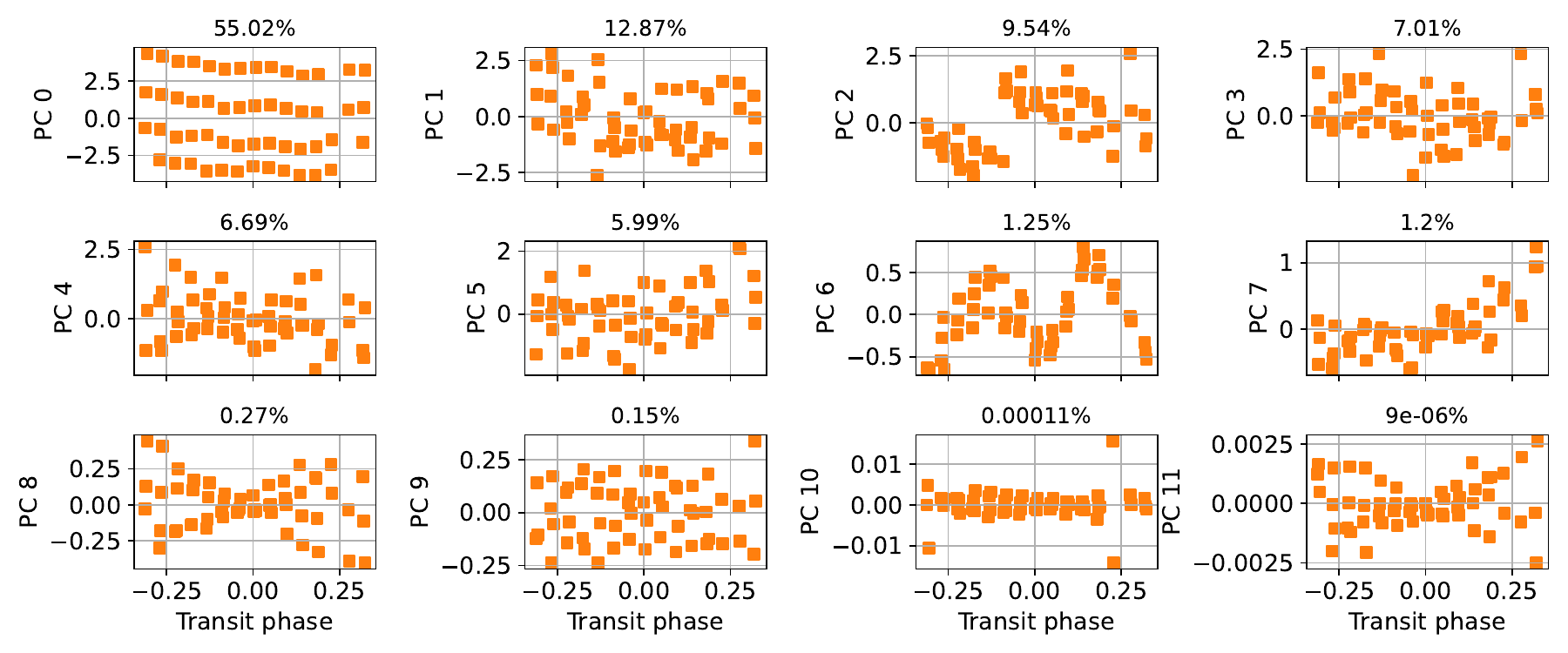}{0.95\textwidth}{}}
    
    \vspace{-0.5cm}
    
     \caption{\textbf{Top:} The principal components for Visit 1. \textbf{Bottom: }The principal components for Visit 2, both plotted against the KELT-9b orbital phase. Each PC's variance is shown in the respective plot title.}\label{fig:pcas}
     
\end{figure*}

PCA transforms a set of correlated or potentially correlated variables into an orthogonal set of uncorrelated variables, each called a principal component. It does so by scaling each variable to unit variance, calculating the covariance matrix of the unit-scaled set, and then decomposes the covariance matrix into eigenvectors, or the principal components (PCs). Each PC is then responsible for some level of variation in the whole data set. We used the python \texttt{scikit-learn} package \citep{scikit-learn} to carry out the PCA, and the PCs for visits 1 and 2 are shown in Figure \ref{fig:pcas}, plotted against the transit phase.

From Figure \ref{fig:pcas}, it is evident how different the $CUTE$ spacecraft states are between Visits 1 and 2. For both visits, PC 0 is likely heavily influenced by the detector temperature, as each spacecraft orbit, which typically contains 4 exposures (though there are a few in each visit with only 3 or 2 exposures) begins with a higher temperature as the spacecraft enters the Earth's shadow, with successively lower temperatures as the spacecraft cools within the shadow. Hence, the appearance of 4 distinct rows in PC 0 for both Visits 1 and 2 is due to exposures being taken at similar locations in a given orbit. There are PCs that exhibit dips near the mid-transit time, which may have otherwise been mistaken as a transit signal. For example a dip is very evident in PC 6 for Visit 2, and more subtly in PC 0 for Visit 1. We emphasize that no $CUTE$ observations were used in the PCA for either visit, only CCD temperature and spacecraft orientation values.

We model the total flux in an exposure, $f$, as the sum of a stellar flux offset, F$_{0}$, a transit light curve model in both time $t$ and transit parameters $\Theta$, $T(t, \Theta)$, and a systematics model consisting of the sum of first-order polynomials for the 12 PCs, additional first-order polynomials for the jitter components, $j$, along the $x$, $y$, and $z$ spacecraft axes:

\begin{equation}
    f = F_{0} + T(t,\Theta) + \sum^{i \leq 12}_{i=1} a_{i}PC_{i} + \sum^{z}_{i=x} b_{i}j_{i}
\end{equation}\label{eq:flux_mod}

\noindent where a$_{i}$ are the coefficients for the PCs and $b_{i}$ are the coefficients for the jitter term $j$. As shown in Figure \ref{fig:pcas}, each PC does not contribute equally to the total variance in the PC dataset, but rather each subsequent PC has a small total variance contribution. PC 0 has the largest variance out of all the PCs, with 49.61\% for Visit 1 and 55.02\% for Visit 2. 

\input{k9_fixed_params}

The use of PCA with the chosen parameters limits us to a first-order polynomial. The raw $CUTE$ counts correlate linearly with the CCD temperature, and since PCA produces a set of orthogonal components, one of which is strongly linear with $CUTE$ data, all other PCs are limited to a first-order polynomials as well.

We use \texttt{batman} \citep{Kreidberg2015_batman} as $T(t, \Theta)$ and let only the planetary to stellar radius ratio, \radrat\, float as a free parameter. The planet's mid-transit time, semi-major axis, eccentricity, longitude of periapsis, and orbital inclination were fixed to nominal values listed in Table \ref{tab:k9_fixed_params}. The $CUTE$ data does not have enough coverage during ingress and egress to fit for limb-darkening coefficients, so we instead used \texttt{limb-darkening} package \citep{Espinoza2015} to calculate quadratic-law limb-darkening coefficients using stellar parameters from Table \ref{tab:k9_fixed_params}. The \texttt{lmfit} python package was used to carry out an MCMC fit to each model.\footnote{We provide details of this process in Appendix \ref{append:lmfit}.}

We first ran Equation \ref{eq:flux_mod} without including jitter to identify which PCs were necessary to produce the best fit. To identify a starting point for which PCs should be included in the model, we use the KELT-9b optical transit depth, $\sim$0.8\%, to define the minimum variance a PC must have to necessarily be included in the model. This requires that PCs 0 -- 8 are included for Visit 1 and PCs 0 -- 9 for Visit 2. We then reran the model including successive PCs until all 12 were included. As is done in \cite{Cubillos2023}, \cite{Gressier2023}, and \cite{Sreejith2023}, the best-fit model is identified as that which minimizes the corrected Akaike information criterion for small samples, AICc, defined as 

\begin{equation}\label{eq:aic}
    AICc = \chi^{2} + \frac{2k(k+1)}{n-k-1}
\end{equation}

\noindent where $k$ is the number of free parameters and $n$ are the number of data points. Once the best-fit model is found for the NUV broadband light curve, the same functional form is used to fit the spectral light curve. An analysis of the background levels across the spectral axis of the CCD show variation, indicating that the systematics model will vary as a function of pixel location along the spatial axis on the detector. Additionally, scattered light levels across the CCD are not flat (e.g. see Figure 8 in \cite{Egan2023}). Therefore, we expect the same general trends to appear across $CUTE$'s bandpass, but with different magnitudes that are dependent on the spatially-inhomogeneous scattered light levels. This is in contrast to the assumption that systematics vary weakly with wavelength, as is commonly found in studies of $HST$ data (e.g. \citealt{Wakeford2016, Sing2019, Gressier2023}).



%% file: k9_fixed_params.tex
\begin{deluxetable*}{ccccc}




\tablecaption{KELT-9 system parameters, $\Theta$\label{tab:k9_fixed_params}}


\tablehead{\colhead{Parameter} & \colhead{Unit} & \colhead{Symbol} & \colhead{Value} & \colhead{Source}
} 
\startdata
Stellar Effective Temperature  &  K     &  T$_{eff}$      &  10170      &  \cite{Gaudi2017}   \\
Stellar Mass & M$_{\odot}$ & M\sr & 1.978 $\pm$ 0.023 &  \cite{Hoeijmakers2019} \\
Stellar Radius & R$_{\odot}$ & \rs & 2.178 $\pm$  0.011 & \cite{Hoeijmakers2019} \\
Stellar Surface Gravity  &  cgs        &  log(g)         &  4.093      &  \cite{Gaudi2017}   \\
Semi-major axis         &  AU   &  a      &  0.03368    &  \cite{Borsa2019}   \\
Planet Mass & \mj & \mpj & 2.44 $\pm$ 0.70 & \cite{Hoeijmakers2019} \\
Planet Radius & \rj & R$_{p}$ & 1.783 $\pm$ 0.009 & \cite{Hoeijmakers2019} \\
Orbital Period  &  Days         &  P      &  1.48111871  &  \cite{Ivshina2022}   \\
Transit Center Time & Julian Date      &   T$_{c}$ & 2457095.68572 &  \cite{Ivshina2022}  \\
Inclination     &  degree       &  $i$    &  86.79      &  \cite{Gaudi2017}   \\
Eccentricity    &  degree       &  $\varepsilon$  &  0  &  \cite{Jones2022}   \\
Argument of Periastron  &  degree       &  $\varpi$       &  90         &  \cite{Jones2022}   \\
\enddata




\end{deluxetable*}

%% file: results.tex
\section{Results \& Discussion}\label{sec:results}

We show the broadband NUV light curves for Visits 1 and 2 in Figure \ref{fig:white_light_v1v2}; the top plots show the uncorrected counts normalized to the first three out-of-transit orbits, the middle plot shows the best fit broadband NUV light curve atop the systematics-removed $CUTE$ data with the error regions shaded in the respective colors, and the bottom plot shows the residuals. For both visits, the best model included the minimum number of PCs; when each additional PC was included in the model, the AICc grew by less than 1 for both visits, indicating that additional PCs do not improve the fit and are functionally the same. For both visits, the best-fit model was found when only the x-axis jitter was included. We find this reasonable as jitter along the x-axis translates into jitter along the shorter telescope axis, or along the spatial direction of the detector (i.e. vertically as shown in Figure \ref{fig:1d2d}).

For both visits, Figure \ref{fig:white_light_v1v2} demonstrates the importance of a long out of transit baseline for constraining the out-of-transit continuum level in the presence of noisy data. Whereas $CUTE$ was able to observe KELT-9b for approximately 22 hours for each transit, $HST$ STIS using the MAMA detectors, the other operating instrument capable of obtaining NUV transmission spectroscopy, is typically limited to a maximum of 5 orbits per target, or approximately 8 hours, with rare exceptions being granted to allow up to 6 orbits \citep{STIS_C32_handbook}.

\begin{figure*}[t!]
    \gridline{\fig{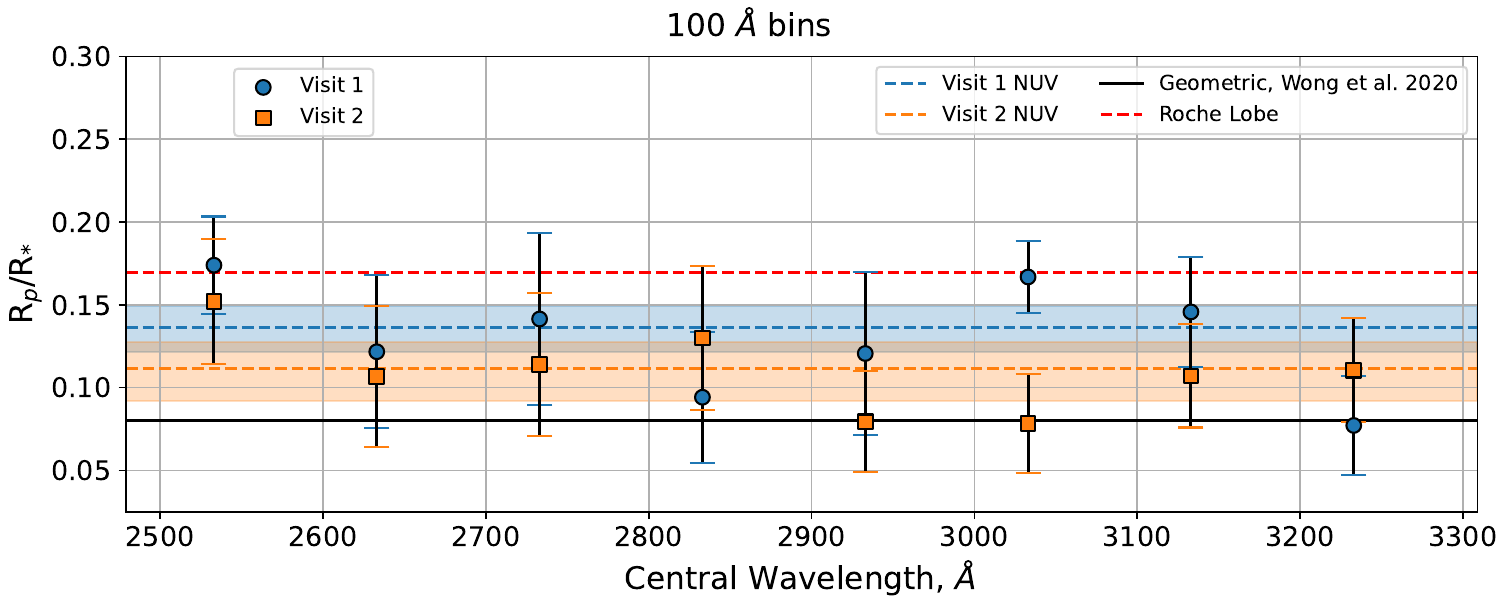}{0.95\textwidth}{}}
    \vspace{-0.8cm}
    \gridline{\fig{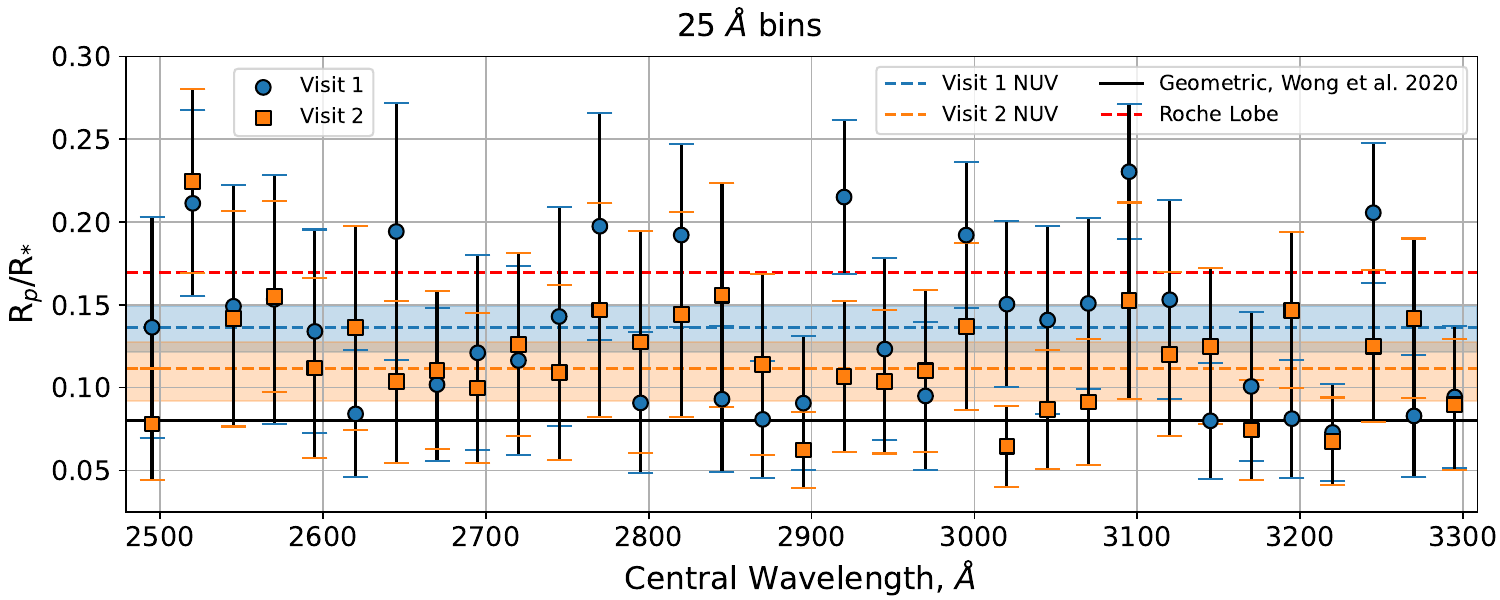}{0.95\textwidth}{}}
    
    \vspace{-0.5cm}
    
     \caption{\textbf{Top:} Transmission spectrum for Visit 1 and Visit 2 with 100 \AA\ wide bins, corresponding to R $\sim$ 28. \textbf{Bottom: }Transmission spectrum for Visit 1 and Visit 2 with 25 \AA\ wide bins, corresponding to R $\sim$ 112. On both plots, the Roche Lobe radius is shown as the red dashed line, and the visible light \radrat\ from  \cite{Wong2020} is shown as the black solid line.} \label{fig:100_transspect}
\end{figure*}

While the broadband raw light curves for Visits 1 and 2 vary considerably from each other, the best fit light curves are consistent with each other. The best fit \radrat\ for Visits 1 and 2 respectively are: \radrat$_{V1}$ = 0.136\lowhi{0.0146}{0.0125} with a reduced chi-squared $\chi^{2}_{\nu}$ = 1.0053, and \radrat$_{V2}$ $=$ 0.111\lowhi{0.0190}{0.0162} with $\chi^{2}_{\nu}$ = 0.9987. Compared to the $TESS$ red-optical \radrat$_{opt}$ $=$ 0.0804 from \cite{Wong2020}, KELT-9b appears an average of 1.54$\times$ larger in the NUV. This means that the NUV broadband transit is probing relatively low pressures in the atmosphere. The corresponding Roche lobe filling factor is larger than that observed for other hot Jupiters such as HD 209458b and HD 189733b but not as large as for the ultra-hot Jupiter WASP-121b (e.g., compare Figure \ref{fig:100_transspect} in this paper to Figure 8 in \citealt{Cubillos2023}). In principle, the results for KELT-9b here appear similar to those for WASP-189b where $CUTE$ observations indicated that the upper atmosphere of the planet is hotter and more extended than expected (\citealt{Sreejith2023}). The interpretation of the observations of KELT-9b, however, is complicated by the uncertainty on planet mass that allows for a range of solutions on atmospheric structure and mass loss (\citealt{Borsa2019, Hoeijmakers2019, Lowson2023}).

\begin{figure*}
    \gridline{\fig{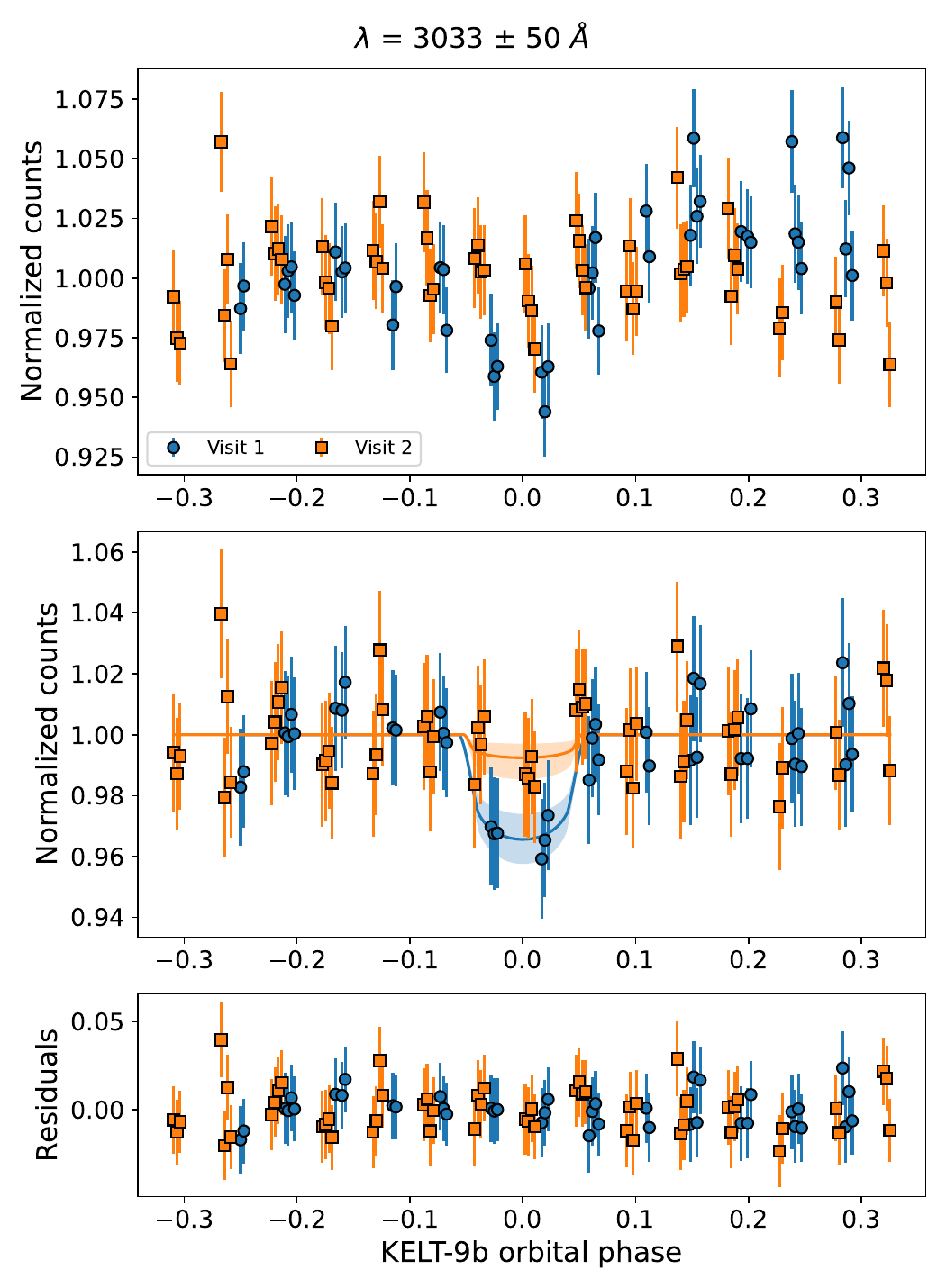}{0.45\textwidth}{}
             {\fig{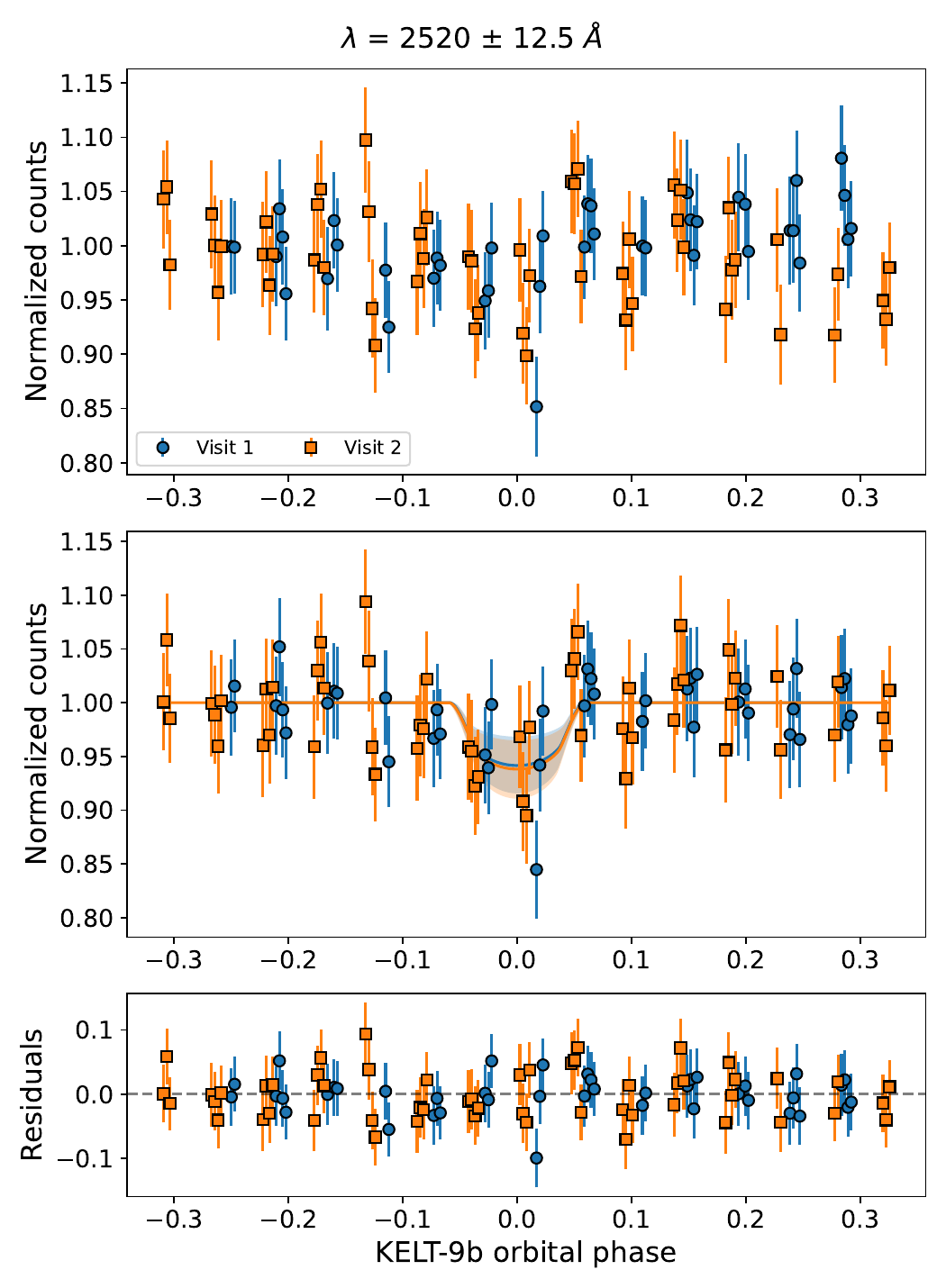}{0.45\textwidth}{}}}
    \vspace{-0.8cm}
    
     \caption{\textbf{Left:} Light curves for Visits 1 and 2 for a 100 \AA\ wide bin centered at 3033 \AA. \textbf{Right: }Light curves for Visits 1 and 2 for a 25 \AA\ wide bin centered at 2520 \AA.} \label{fig:spec_lcs}
\end{figure*}

We attempted a joint fit between the two visits by keeping the \radrat\ as a shared parameter and the systematics models separate for each visit. However, the joint fit produced a \radrat\ equal to the average of the individual fits with a $\chi^{2}_{\nu}$ = 1.35. We therefore keep the remaining analysis to a per-visit basis.

To further explore the light curve differences, we produced a transmission spectrum with 100 \AA\ and 25 \AA\ wide bins, corresponding to R $\sim$ 28 and 112 respectively, for each visit. Using the approximation from \cite{Eggleton1983} and the values in Table \ref{tab:k9_fixed_params}, we calculate the Roche Lobe radius to be R$_{L}$ = 2.017 \rp \ and include it as a visual reference. For each bin in Figure \ref{fig:100_transspect}, we used the same fitting procedure as in Section \ref{sec:dataAnalysis}. These transmission spectra are shown in Figure \ref{fig:100_transspect} and shown in Tables \ref{tab:k9_rprs} and \ref{tab:k9_rprs_25} respectively.
\input{rprs_table}
\input{rprs_25_table}

In the 100 \AA\ transmission spectrum, the two visits are generally consistent with each other except at the central wavelength of 3033 \AA, where Visit 1 exhibits a transit depth about 2.5$\sigma$ times greater than Visit 2 and is $\geq$3$\sigma$ larger than the optical light measurement from \cite{Wong2020}. The light curves for that bin is shown in Figure \ref{fig:spec_lcs}.

A transmission spectrum with smaller bins would aid in assessing what might be responsible for the difference between the two visits at 3033 \AA\ and to search for signs of strongly absorbing ions. Additional transit observations would be necessary to increase the required signal to noise ratios for finer line detection. The 25 \AA\ spectrum, shown in the bottom half of Figure \ref{fig:100_transspect}, indicates that Visits 1 and 2 are generally consistent with each other within their 1$\sigma$ uncertainties. The shape of the 25 \AA\ transmission spectrum is generally more noisy for Visit 1 than it is for Visit 2, which may be due to unidentified or remaining systematics.

There are a few hints of extended absorption in the $CUTE$ NUV transmission spectrum. The 25 \AA\ bin centered at 2520 \AA\ shows that both visits are $>$ 1$\sigma$ above the broadband visible \radrat\, with Visit 1 being 1.3$\sigma$ greater and Visit 2 being 2.05$\sigma$ greater. Three additional 25 \AA\ bins in Visit 1 also show $>$1$\sigma$ above the broadband NUV \radrat, located at 2920 \AA\, 3095 \AA\, and 3245 \AA. The bin at 3095 \AA\ lies $>$1$\sigma$ above the planet's Roche lobe, directly suggesting an escaping atmosphere. However, without small wavelength bins, it is challenging to discern which atomic species might be responsible for the increased absorption.

As discussed in several papers, the NUV is filled with neutral and ionized atoms that can induce extended transit depths, and KELT-9b has had a wealth of atomic detections made in optical wavelengths, motivating a search for them in the NUV where atomic transitions tend to be stronger than in the optical. For example, Fe, which has several lines present in $CUTE$'s bandpass (e.g. \citealt{Sing2019}), has been detected several times with ground-based spectrographs: \cite{Cauley2019} detected 9 individual lines of \ion{Fe}{2} with the PEPSI spectrograph; both \ion{Fe}{1} and \ion{Fe}{2} were observed in \cite{Lowson2023} with TRES; and \ion{Fe}{2} definitively detected in \cite{Borsato2023FOCES} with FOCES and in \cite{BelloArufe2022} with FIES. We note that the spectrum of WASP-121b that shows the clearest NUV signatures of escaping metals to date (\citealt{Sing2019}) can be explained by \ion{Fe}{2} lines that dominate the spectrum and the strong \ion{Mg}{2} h\&k lines (see Figure 23 in \citealt{Huang2023} for a model fit). While the absorption in the $CUTE$ spectrum at 2520 \AA\ coincides with \ion{Fe}{2} lines, absorption due to \ion{Fe}{2} at 2600 \AA, which was seen in WASP-121b \citep{Sing2019}, is not present in the $CUTE$ spectrum. At the same time, neither the model or observations of \ion{Fe}{2} in WASP-121b show strong features at 2920 \AA\ or 3095 \AA. 

The \ion{Mg}{2} resonance lines at 2800 \AA\ are expected to be present in KELT-9b as they have been observed in the ultra-hot Jupiters WASP-12b \citep{Fossati2010, Haswell2012}, WASP-189b \citep{Sreejith2023} and WASP-121b \citep{Sing2019}. Interestingly, there is no suggestion in the 100 \AA\ or 25 \AA\ transmission spectra of significantly extended absorption around 2800 \AA. It is possible that residual systematics are masking the transit signal, however. In any case, the available data are unable to produce meaningfully smaller resolution transmission spectra.

%% file: rprs_table.tex
\begin{deluxetable}{ccc}




\tablecaption{KELT-9b Transmission Spectrum, 100 \AA\ bin widths\label{tab:k9_rprs}
}


\tablehead{\colhead{Central $\lambda$, \AA} & \colhead{Visit 1 \radrat} & \colhead{Visit 2 \radrat}
} 
\startdata
2533 & 0.174\lowhi{0.0294}{0.0246} & 0.152\lowhi{0.0378}{0.0300} \\
2633 & 0.122\lowhi{0.0461}{0.0366} & 0.107\lowhi{0.0427}{0.0361} \\
2733 & 0.141\lowhi{0.0521}{0.0399} & 0.114\lowhi{0.0433}{0.0362} \\
2833 & 0.094\lowhi{0.0394}{0.0363} & 0.130\lowhi{0.0435}{0.0346} \\
2933 & 0.121\lowhi{0.0493}{0.0410} & 0.079\lowhi{0.0305}{0.0301} \\
3033 & 0.167\lowhi{0.0218}{0.0185} & 0.078\lowhi{0.0299}{0.0306} \\
3133 & 0.146\lowhi{0.0333}{0.0247} & 0.107\lowhi{0.0311}{0.0249} \\
3233 & 0.077\lowhi{0.0299}{0.0311} & 0.111\lowhi{0.0314}{0.0240} \\
\enddata




\end{deluxetable}

%% file: rprs_25_table.tex
\begin{deluxetable}{ccc}




\tablecaption{KELT-9b Transmission Spectrum, 25 \AA\ bin widths\label{tab:k9_rprs_25}}


\tablehead{\colhead{Central $\lambda$, \AA} & \colhead{Visit 1 \radrat} & \colhead{Visit 2 \radrat}
} 
\startdata
2495 & 0.136\lowhi{0.0666}{0.0578} & 0.078\lowhi{0.0335}{0.0440} \\
2520 & 0.211\lowhi{0.0562}{0.0429} & 0.225\lowhi{0.0556}{0.0458} \\
2545 & 0.149\lowhi{0.0732}{0.0702} & 0.142\lowhi{0.0650}{0.0584} \\
2570 & 0.153\lowhi{0.0750}{0.0660} & 0.155\lowhi{0.0576}{0.0475} \\
2595 & 0.134\lowhi{0.0616}{0.0548} & 0.112\lowhi{0.0545}{0.0624} \\
2620 & 0.084\lowhi{0.0383}{0.0575} & 0.136\lowhi{0.0615}{0.0583} \\
2645 & 0.194\lowhi{0.0775}{0.0572} & 0.104\lowhi{0.0488}{0.0572} \\
2670 & 0.102\lowhi{0.0463}{0.0506} & 0.110\lowhi{0.0477}{0.0414} \\
2695 & 0.121\lowhi{0.0588}{0.0632} & 0.100\lowhi{0.0454}{0.0475} \\
2720 & 0.116\lowhi{0.0571}{0.0611} & 0.126\lowhi{0.0554}{0.0504} \\
2745 & 0.143\lowhi{0.0660}{0.0545} & 0.109\lowhi{0.0528}{0.0568} \\
2770 & 0.197\lowhi{0.0685}{0.0512} & 0.147\lowhi{0.0644}{0.0525} \\
2795 & 0.091\lowhi{0.0426}{0.0569} & 0.128\lowhi{0.0671}{0.0638} \\
2820 & 0.192\lowhi{0.0553}{0.0411} & 0.144\lowhi{0.0618}{0.0502} \\
2845 & 0.093\lowhi{0.0442}{0.0508} & 0.156\lowhi{0.0675}{0.0591} \\
2870 & 0.081\lowhi{0.0353}{0.0491} & 0.114\lowhi{0.0546}{0.0569} \\
2895 & 0.091\lowhi{0.0405}{0.0505} & 0.062\lowhi{0.0231}{0.0355} \\
2920 & 0.215\lowhi{0.0467}{0.0376} & 0.107\lowhi{0.0456}{0.0434} \\
2945 & 0.123\lowhi{0.0550}{0.0466} & 0.104\lowhi{0.0434}{0.0398} \\
2970 & 0.095\lowhi{0.0445}{0.0599} & 0.110\lowhi{0.0488}{0.0459} \\
2995 & 0.192\lowhi{0.0441}{0.0354} & 0.137\lowhi{0.0503}{0.0390} \\
3020 & 0.150\lowhi{0.0501}{0.0372} & 0.065\lowhi{0.0244}{0.0364} \\
3045 & 0.141\lowhi{0.0566}{0.0446} & 0.087\lowhi{0.0360}{0.0346} \\
3070 & 0.151\lowhi{0.0516}{0.0399} & 0.091\lowhi{0.0380}{0.0371} \\
3095 & 0.230\lowhi{0.0409}{0.0328} & 0.153\lowhi{0.0593}{0.0461} \\
3120 & 0.153\lowhi{0.0600}{0.0455} & 0.120\lowhi{0.0496}{0.0398} \\
3145 & 0.080\lowhi{0.0350}{0.0471} & 0.125\lowhi{0.0470}{0.0368} \\
3170 & 0.101\lowhi{0.0449}{0.0450} & 0.074\lowhi{0.0304}{0.0369} \\
3195 & 0.081\lowhi{0.0356}{0.0483} & 0.147\lowhi{0.0470}{0.0364} \\
3220 & 0.073\lowhi{0.0294}{0.0415} & 0.067\lowhi{0.0265}{0.0403} \\
3245 & 0.206\lowhi{0.0422}{0.0337} & 0.125\lowhi{0.0461}{0.0366} \\
3270 & 0.083\lowhi{0.0370}{0.0520} & 0.142\lowhi{0.0483}{0.0383} \\
3295 & 0.094\lowhi{0.0427}{0.0485} & 0.090\lowhi{0.0396}{0.0473} \\
\enddata




\end{deluxetable}

%% file: conclusion.tex
\section{Conclusion}\label{sec:conclusion}

Herein we presented NUV transmission spectroscopy of the ultra-hot Jupiter KELT-9b obtained with the $CUTE$ CubeSat. Two consecutive transit observations of KELT-9b, made on September 28th and 29th 2022, show differing raw light curves but consistent best-fit NUV radii. In order to maximize the number of spacecraft parameters used to detrend the data, we used principal component analysis to transform a set of correlated spacecraft parameters into a set of orthogonal parameters. From that, the two NUV broadband light curves produced consistent transit depths, with Visit 1 having \radrat = 0.136\lowhi{0.0146}{0.0125} and Visit 2 having \radrat $=$ 0.111\lowhi{0.0190}{0.0162}.

We further produced two transmission spectra, with 100 \AA\ and 25 \AA\ wide bins respectively. Within the error bars, the transmission spectra are consistent between the two visits. However, the data do not have enough sensitivity to spectroscopically isolate specific atomic absorbers. The 25 \AA\ transmission spectrum contains three bins with absorption above the Roche lobe boundary with greater than 1$\sigma$ significance. However, Visit 2 does not exhibit the same signal, and in general the Visit 1 and Visit 2 25 \AA\ transmission spectra have different shapes, perhaps suggesting that there are uncharacterized systematics remaining in the spectral light curves. It is part of our future work to explore these hints of atmospheric escape.

Despite the data quality, the broadband NUV transit depth of the low-resolution $CUTE$ spectra show promise for the direct detection of an escaping atmosphere on KELT-9b. We are continuing to assess additional methods for removing the systematics present in the data and resolving the disagreement between Visits 1 and 2 in the higher resolution transmission spectra. Additional higher resolution spectra obtained with e.g. $HST$ STIS will likely provide rich insight into KELT-9b's upper atmosphere.

%% file: appendix_lmfit.tex
\section{Using lmfit}\label{append:lmfit}
\texttt{lmfit} enables a modular approach to model creation and curve fitting. The \texttt{Model} class turns a given function into a model to be fit with, and several \texttt{Model} classes with different independent variables and parameters can be added together, producing a \texttt{CompositeModel} class. The \texttt{models} module has several \texttt{Model} classes built in, including a polynomial model, called \texttt{PolynomialModel} up to the 7th degree. In a \texttt{CompositeModel}, a \texttt{Model} class can only have a single independent variable even if instantiated multiple times, $i.e.$ the user cannot instantiate \texttt{PolynomialModel(x)} $+$ \texttt{PolynomialModel(y)} as they have two different independent variables. To enable the sum of polynomials in Eq. \ref{eq:flux_mod}, we utilized the open source nature of \texttt{lmfit} and added several more Polynomial models to account for all PCs and jitter terms (i.e. \texttt{PolynomialModelB}, \texttt{PolynomialModelC}, \texttt{PolynomialModelD}, etc.).